\documentclass[twocolumn]{emulateapj}

\slugcomment{Science and Evolution of Gemini Observatory, 22-26 July 2018, San Francisco, U.S.A.}

\shorttitle{Quasars and Bell's Inequality}
\shortauthors{Steinbring}

\input epsf
\def\plotone#1{\centering \leavevmode
\epsfxsize=1.0\columnwidth \epsfbox{#1}}

\begin{document}

%\begin{center}

\title{Potential for Quantum-Mechanical Tests Using Quasars, \\as Illuminated by Gemini Archival Data}

\author{Eric Steinbring\altaffilmark{1}}

\altaffiltext{1}{National Research Council Canada, Herzberg Astronomy and Astrophysics, Victoria, BC V9E 2E7, Canada}

\begin{abstract}
There has been recent interest in quantum-mechanical tests aided by distant quasars. For two quasars of sufficient redshift at opposite directions on the sky, light-travel-time arguments can assure the acausality of their photons. And if those photons are used to set parameters in an Earth-based apparatus, coincidence cannot be due to their communication, closing the so-called ``freedom of choice" loophole in the experiment.  But this assumes no other interference right up to detection, including correlated instrumental errors, which must be carefully constrained.  The Gemini North and South Multi-Object Spectrograph (GMOS) twins can simultaneously view pairs of quasars up to 180 degrees apart on the sky, and already provide a significant baseline record to investigate this. All GMOS broadband imaging frames were searched to find those that happen to contain a known quasar together with a suitable comparison star. Although individual photometry can be noisy among these $0.1 < z < 6$ sources, in the aggregate, average site conditions and their relative photometric zeropoints are well characterized. The resulting dataset constitutes about 2 million correlated quasar-observation pairs over 14 years.  A preliminary analysis of that is presented, with the intriguing result that paired-flux differences across the whole sky weakly deviate from flatness, to the limit consistent with Bell's Theorem. Can Gemini be used to prove the ``spooky action at a distance" expected of quantum mechanics? Some prospects for future work and a more definitive test are considered.
\end{abstract}


\section{A simple theoretical bound}

A promising route for closing the so-called ``freewill" loophole of laboratory quantum-mechanical tests is to exclude bias by setting experimental parameters via photons from quasars (Friedman et al. 2013; Gallicchio et al. 2014). In the classic benchtop measurement of entangled polarization states, an experimenter could now set the polarimeter orientation based on the photon time of arrival from an external, unrelated source.  Better: two separate, distant sources would each provide photons that cannot have colluded in the result.  Although a test using stars within the Milky Way was recently achieved by Handsteiner et al. (2017), pushing any potential setting ``conspiracy" back hundreds of years, quasars may prove definitive: they are also pointlike, but visible up to high redshift, and combined with large angular separation on the sky can place these entirely outside each others' light cone; for separations of 180 degrees this occurs when both have $z\geq3.6$. The independence of settings triggered by those photons cannot have been spoiled by their communication. But could the fluxes of two quasars be correlated, and so still conspire?

Beyond viewing in opposite directions on the sky, a problem with pursuing this experiment from the ground is that it tests against a correlation with separation angle $\theta$ with form $-\cos{\theta}$ (thin black curve, bottom panel of Figure 1) which just happens to inversely correlate with target airmasses from any single site. Bell's inequality predicts, in the classic test of entangled quantum states, a proportion of correlated arrival-angle occurances $$P(\theta) = |\cos{2\theta} - \cos{\theta}| + \cos{\theta}, \eqno(1)$$ which exceeds unity for all angles less than 90 degrees, thus may be discernable from observational errors, if compared against larger angles. Under clear skies, attenuation is roughly linear with airmass (as are sky brightness and seeing to first order) so for targets at zenith angle $\theta$, and under an airmass of 2, the maximal variation between two sites geographically separated by 90 degrees is $$E(\theta) = \Big{|} 2 - {1\over{\cos{\theta}}} \Big{|}, \eqno(2)$$ where $\alpha$ is half the mean attenuation in magnitudes, with differences $$\Delta{A} = \alpha E, ~~\Delta{B} = \beta E, ~~\Delta{Q} = \gamma E \eqno(3)$$ in extinction, sky-brightness and seeing, and where $\beta$ is in magnitudes per square arcsec and $\gamma$ in arcseconds. Therefore, the maximum angle-dependent difference of $n$ unbiased quasar pairs per angular bin, taken from the same brightness distribution of width $\omega$ and amplified by equation 1, should over a long-term average have $$S/N \sim {{\sqrt{n} \omega P} \over{\alpha + \beta \gamma^2 + \delta}}. \eqno(4)$$ The signal strength amounts to excess relative to flatness, and so the absolute values of differences will serve.

\begin{figure}
\plotone{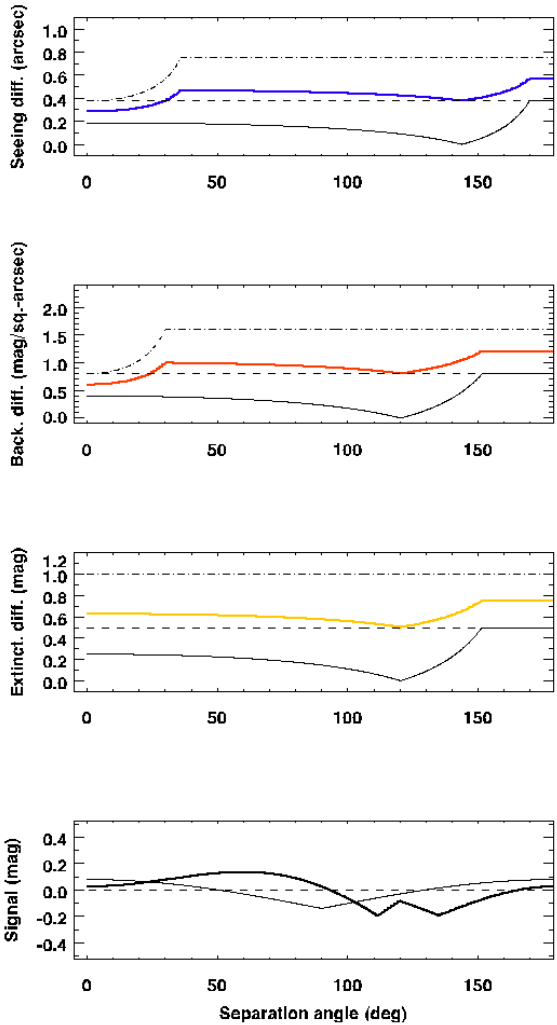}
\caption{Model differences in seeing, sky-brightness, and exinction between any two positions on the sky up to 180 degrees apart, viewed from two different sites, here plotted for Gemini. Instantaneous lower and upper limits are indicated, giving means for coincident observations (thick curves) and long-term averages for combining views from each single site (dashed lines). Below is an expected difference in flux of two quasars selected at random from a brightness distribution of width 1.00 mag, if their fluxes were correlated as per Bell's Theorem, and reaching $S/N=3$ as per equation 4 (thick black curve).}\label{figure_model_differences_with_angle}
\end{figure}

\section{Many serendipitous quasar images}

An ideal test would view two quasars simultaneously with identical instruments, and to achieve angular separations larger than about 120 degrees in the optical from the ground these must be in opposite hemispheres. Thankfully, such a pair of (nearly identical) instruments exists, and has been operating (nearly continuously) for over 15 years.
 
Every archival Gemini Multi-Object Spectrograph (GMOS) imaging frame was searched for instances of a known quasar falling into the field. The Million Quasars Catalog (MILLIQUAS), Version 4.8 of 22 June 2016 was used, providing a redshift, magnitude and a Sloan Digital Sky Survey (SDSS) blue-red color for each target. That was cross-correlated with the full Canadian Astronomy Data Center (CADC) catalog of science frames obtained with GMOS North and South, complete to the end of 2015, using the Common Archive Observation Model (CAOM) Table Access Protocol (TAP) web service and custom Python scripts.

This provided 28374 cases where a quasar was within a GMOS $g$, $r$, $i$ or $z$-band 5 arcminute-wide science field; deleting those corrupted or otherwise unusable yielding 20514 sources with $0.1 < z < 6$ and no noticable dependence on sky position, and a typical exposure time of about 150 seconds. Each of those was also searched for a suitable, nearby unsaturated comparison star from the Fourth U.S. Naval Observatory CCD Astrograph Survey (UCAC4) to serve as a photometric calibration, resulting in a total sample of 9713 usuable frames under good sky conditions, through the full range of right ascension, and between $-79^{\circ}$ and $+82^{\circ}$ declination.

Synthetic aperture photometry was carried out on all objects, employing a 4 arcsec diameter aperture throughout. Each frame was first corrected for detector gain, background subtracted, and then de-extincted using the comparison star relative to its catalog UCAC4 $r$ magnitude and an appropriate zeropoint for the filter. The quasar photometry was also shifted by the mean filter correction to $r$, and each taken as a differential from its catalog magnitude. In this way, all photometry is relative to the sample average of $r=20$ mag, and sky-brightness limited, with some down to the expected 5-$\sigma$ point-source limit of about 23 mag; see Figure 2.

\begin{figure}
\plotone{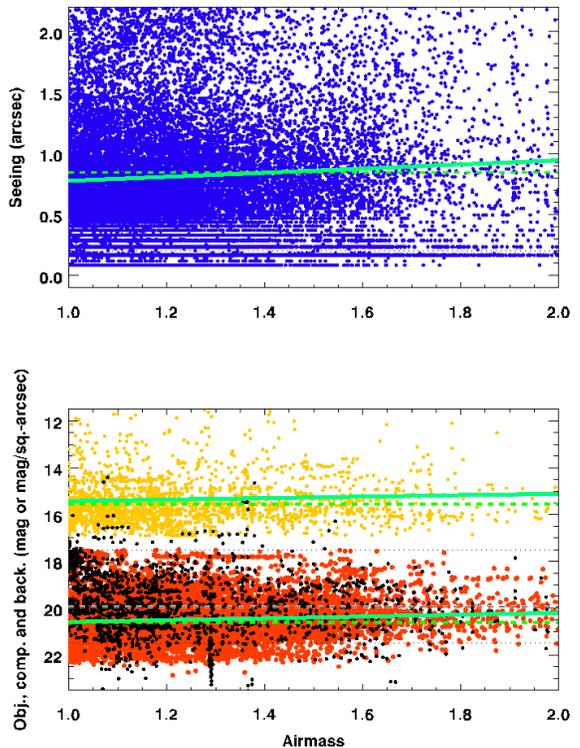}
\caption{All quasar magnitudes obtained (filled black circles) as a function of airmass. Also shown are the UCAC4 $r$-band magnitudes for the comparison stars in each field (yellow) and their background sky brightnesses (red). Above: image quality estimates for each frame (blue); data are spurious below a cutoff at about 0.2 arcsec (dotted line).}\label{figure_object_comparison_background_and_seeing_with_airmass}
\end{figure}

Linear least-squares fits to extinction (comparison magnitudes), sky-brightness and seeing (image quality) are shown in Figure 2 in green (averages: dashed), giving $\alpha\approx 0.25$ mag, $\beta\approx 0.40$ mag per square-arcsec, and $\gamma\approx 0.38$ arcsec. With $\delta=0.25$ mag, this suggests, using equation 4, that the minimum sample per angular bin required to detect the influence of $P$ is on the order of 300.

\section{Some preliminary correlated results}

The UTC time of each frame was compared to those previous to find overlapping exposures, while demanding that catalog SDSS colors were within 0.2 magnitudes of each other, less than the typical stellar photometric error and color corrections between filters. It was hoped that there would be many coincident cases (North and South together) but significant overlaps with good data occured only when two quasars were found within the same frame, occuring 310 times (light blue: 0 degrees separation in Figure 3).

\begin{figure}
\plotone{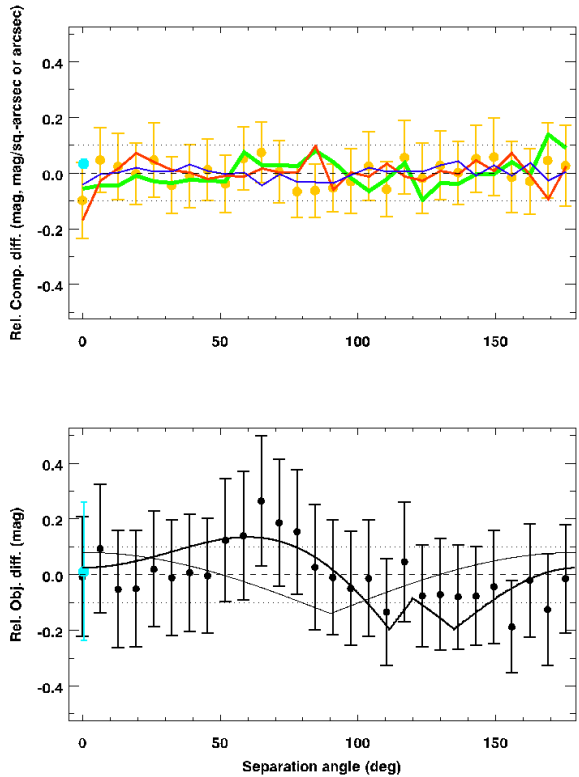}
\caption{Observed quasar-magnitude differences versus separation angle, averaged in 6-degree-wide bins. The light-blue filled circle is for the coincident cases only.  Error bars are standard deviations in each angular bin.  Above are the same, for differences in the comparison stars (yellow); the standard deviation of the 2 million cross-correlated differences is 0.02 magnitudes. Compared to scatter, the differences in quasar catalog magnitudes are flatter (green), as are sky brightness (red) and image quality (blue). So those do not seem to explain the apparent curve as per the model for Bell's Theorem (black). Results are similar, but with expectedly more scatter, if only $r$-band frames are used (not shown). Yes, the fluxes of quasars here do indeed seem ``spookily" correlated.}\label{figure_differences_with_angle_final}
\end{figure}

\section{Summary and possible future work}

A clean quantum-mechanical experiment - excluding collusion between parameter settings via photons from distant quasars - assumes the fluxes of those objects are not correlated.  Fair, random sampling across the sky should be able to falsify that, and some preliminary results of such a test is presented (from Steinbring, 2018, {\it in preparation}). Almost 10000 sufficiently deep quasar images were found for Gemini GMOS North and South, complete with a nearby unsaturated UCAC4 star, allowing photometry with a global zeropoint uncertainty of about 2\% over a sample spanning 14.3 years. These are just sufficient to show a lack of flatness in their relative quasar flux-differences with angular separation that is intriguingly consistent with Bell's inequality. But the dataset provides little information on how those individual quasars (or the calibration stars) may have varied during that time. Another confusing factor might be shifting bandpasses with redshift, and correcting to a common color; a better result may be obtained spectroscopically, focussed on bright emission lines. Although this dataset provides only a small number of coincident observations, yielding just those cases where two quasars were in the same field, that could prove a more robust path, and is possible for separations up to 180 degrees with Gemini.

\begin{acknowledgements}
This research used the facilities of the CADC operated by the NRC of Canada with the support of the Canadian Space Agency. I thank David Bohlender in particular for his kind assistance with TAP scripts.
\end{acknowledgements}

\end{document}